\documentclass{aa}
\usepackage{graphicx}
\begin{document}
\title{Study of molecular layers in the atmosphere of the supergiant star $\mu$~Cep by interferometry in the K band\thanks{Based on observations collected at the IOTA 
interferometer, Whipple Observatory, Mount Hopkins, Arizona.}}

\titlerunning{Molecular layers in the atmosphere of $\mu$~Cep}

\author{G. Perrin \inst{1}
\and S.T. Ridgway \inst{1}\fnmsep \inst{2} 
\and T. Verhoelst \inst{1}\fnmsep \inst{3}
\and P.A.  Schuller \inst{4}
\and V. Coud\'e du Foresto \inst{1}
\and W.A. Traub \inst{4}
\and R. Millan-Gabet \inst{5}
\and M.G. Lacasse \inst{4}
}

   \offprints{G. Perrin, \email{guy.perrin@obspm.fr}}

\institute{Observatoire de Paris, LESIA, UMR 8109, F-92190 Meudon, 
France \and National Optical Astronomy Observatories, Tucson, AZ 
85726-6732, USA \and  Instituut voor Sterrenkunde, K.U. Leuven, 
Belgium 
\and Harvard-Smithsonian Center for Astrophysics, Cambridge, MA 
02138, USA
\and Caltech/Michelson Science Center, Pasadena, CA 91125, USA}

   \date{Received ; accepted }


   \abstract{Infrared interferometry of supergiant and Mira stars has recently been reinterpreted as revealing the presence of deep molecular layers.  Empirical models for a photosphere surrounded by a simple molecular layer or envelope have led to a consistent interpretation of previously inconsistent data.  The stellar photospheres are found to be smaller than previously understood, and the molecular layer is much higher and denser than predicted by hydrostatic equilibrium.  However, the analysis was based on spatial observations with medium-band optical filters, which mixed the visibilities of different spatial structures.   This paper reports spatial interferometry with narrow spectral bands, isolating near-continuum and strong molecular features, obtained for the supergiant $\mu$~Cep.  The measurements confirm strong variation of apparent diameter across the K-band.  A layer model shows that a stellar photosphere of angular diameter $14.11\pm0.60$~mas is surrounded by a molecular layer of diameter $18.56\pm0.26$~mas, with an optical thickness varying from nearly zero at 2.15 $\mu$m to $>1$ at 2.39 $\mu$m.  Although  $\mu$~Cep and  $\alpha$~Ori have a similar spectral type, interferometry shows that they differ in their radiative properties. Comparison with previous broad-band measurements shows the importance of narrow spectral bands.  The molecular layer or envelope appears to be a common feature of cool supergiants.

     \keywords{ techniques: interferometric -- stars: fundamental parameters  -- stars: mass-loss -- infrared: stars -- stars: individual: $\mu$~Cep
  }
   }

   \maketitle

\section{Introduction}
Contrary to less massive stars at the same stage of their evolution, late-type supergiant stars are only slightly variable. Nevertheless, they have complex and extended atmospheres which make the determination of their fundamental parameters problematic. Near-infrared windows provide the opportunity to see deeper in the atmosphere and reach the photosphere. 

In a recent paper (Perrin et al. 2004a), we have shown that the near-infrared structure of Betelgeuse, of type M1-2Ia-Iab, can be described by two main components: a photosphere of temperature $3690\pm50$~K and a shell, possibly containing gases such as CO, H$_2$O and SiO, at a temperature of $2055\pm25$~K and located $0.33~R_{\star}$ above the photosphere. Strong water-vapor bands had been detected in $\mu$~Cep and Betelgeuse by \cite{danielson1965} but no explanation was available by that time. The star+layer model has been shown to consistently explain our measurements and those of the  ISI interferometer at $11.15$~$\mu$m of \cite{weiner2000} once the contribution of dust has been removed.  This analysis had been first successfully applied to Mira stars (Perrin et al. 2004b) and led to a consistent understanding of seemingly contradictory measurements from the visible to the near-infrared. Other recent modeling studies of supergiants (Ohnaka 2004a) and Mira stars (Ohnaka 2004b, Weiner 2004) have shown that the near- and mid-infrared H$_{2}$O spectra and the apparent near- and mid-infrared angular sizes of both stellar types can be consistently understood in terms of a star with an envelope of warm H$_{2}$O vapor.  These results altogether confirm and vindicate Tsuji's (Tsuji 2000a, 2000b) reinterpretation of older data (Wing \& Spinrad 1970) as evidence for a $1500\pm500$~K water vapor layer around Betelgeuse and $\mu$~Cep. He located this layer a stellar radius above the photosphere.  He found that more water was required in the case of $\mu$~Cep compared to Betelgeuse, although the two stars have similar spectral types (M2~Iae for $\mu$~Cep). Hydrodynamical models of Mira stars appear capable of producing a high molecular layer in some scenarios (e.g. Tej et al. 2003). This may be understood as due to pulsational energy which offers a mechanism for levitating the upper atmosphere. In supergiants, with lower amplitude atmospheric motions, this energy source may not suffice.

The near-IR spatial interferometry reported heretofore for supergiants was based on medium-band filters.  Since these filters included regions of near-continuum and of strong molecular absorption, the data were not optimally sensitive to the model parameters.  The new observations reported in this paper employ narrow filters in molecular and in continuum bands and reveal directly the wavelength dependence of the apparent stellar diameter and strongly constrain the model parameters.  The near-continuum filters, in particular, offer the clearest view and the cleanest measurement yet possible of the stellar surface.  We will present these observations in the next section and explain the data reduction procedure.  We will model the data with a photosphere and a spherical layer in Sect.~\ref{sec:model}. We will discuss the effective temperature, distance and luminosity of $\mu$~Cep in Sect.~\ref{sec:parameters}. Finally, our results will be compared to other interferometric and to spectroscopic results in Sect.~\ref{sec:discussion}.


\section{Observations and data reduction}
$\mu$~Cep was observed along with other sources, mainly Mira stars, in May 2002. The Mira observations have been published in a separate paper in which the observation and reduction details are described (Perrin et al. 2004b). 

Observations took place between May 31 and June 6, 2002 at the  IOTA 
(Infrared-Optical Telescope Array) interferometer located at the 
Smithsonian Institution's Whipple Observatory on Mount Hopkins, 
Arizona (Traub 1998).  Several baselines of IOTA were used to 
sample visibilities at different spatial frequencies.  The data have 
been acquired with FLUOR (Fiber Linked Unit for Optical Recombination) 
in the version described by \cite{foresto98}. We used the NICMOS3 array developed by \cite{millan-gabet1999} to detect the signals.

Observations were carried out in narrow bands with filters specially 
designed for the observation of evolved stars.  We have used the four filters whose 
characteristics are defined in \cite{perrin2004b} and which are named 
K203, K215, K222 and K239 where the three digits characterize the 
central wavelength.  The two continuum filters, K215 and K222, sample 
the maximum transmission region of the K band, and are spaced well away from the familiar strong molecular bands of CO and H$_{2}$O which occur to the blue and to the red of these filters.  The K203 (H$_{2}$O 
bands) and K239 (H$_{2}$O and CO bands) are located near the edges of the K band 
where stellar flux is attenuated by the poorer transmission of the Earth's
atmosphere due to the absorption by water vapor.

Most observations of $\mu$~Cep were bracketed by 
observations of calibrators.  The characteristics of the calibrators are 
listed in Table~\ref{tab:cal}.  HR~5512 is the latest calibrator with a spectral type of M5~III. It was used only for the May 31, 2002 observations. Comparison of this calibrator with others did not show discrepancies or biases. The effects reported in this paper cannot be attributed to 
calibrators and are independent of which calibrator has been used.  \\

\begin{table}[htbp]
      \caption[]{Reference sources.}
         \label{tab:cal}
         \begin{tabular}{lccc}
            \hline
            \hline
            \noalign{\smallskip}
            HD number & Source  &  Spectral type & Uniform disk  \\
	                 & name &                & diameter (mas)\\
            \noalign{\smallskip}
            \hline
   HD130144 & HR 5512        & M5 III    & $8.28\pm0.41$$^{\mathrm{b}}$\\
	    HD140573 & $\alpha$~Ser     & K2 IIIb    & $4.79\pm0.53$$^{\mathrm{a}}$\\
	    HD146051 & $\delta$~Oph     & M0.5 III  & $9.73\pm0.10$$^{\mathrm{a}}$\\
	    HD187076 & $\delta$~Sge     & M2 II       & $8.73\pm0.44$$^{\mathrm{b}}$\\
	    HD197989 & $\epsilon$~Cyg & K0 III      & $4.40\pm0.22$$^{\mathrm{b}}$\\
	    HD198149 & $\eta$~Cep        & K0 IV      & $2.65\pm0.13$$^{\mathrm{b}}$\\
            \noalign{\smallskip}
            \noalign{\smallskip}
            \hline
         \end{tabular}
\begin{list}{}{}
\item[$^{\mathrm{a}}\,\,$\cite{cohen96}]
\item[$^{\mathrm{b}}$ Photometric estimate]
\end{list}
   \end{table}

The log of the observations is given in Table~\ref{tab:obs}. The 
spatial frequency vector is given in polar coordinates comprising the 
azimuth and the modulus listed under Spatial Frequency.  Fringe 
contrasts have been derived with the procedure explained in 
\cite{foresto97}.  The bias in fringe contrast estimates due to photon 
noise has been removed following \cite{perrin2003a}. Finally, the $\mu$~Cep visibilities have been calibrated using the calibrator measurements according to the procedure described in \cite{perrin2003b}.
   
Fig.~\ref{fig:mu_Cep} shows the observed visibility points.  It is immediately clear that the apparent diameter of the star varies with wavelength.  The two near-continuum filters give the highest visibilities, corresponding to the smallest diameter - consistent with the qualitative expectation that the contribution function will reach greater depths in spectral regions of lower opacity than in regions of higher opacity.  The two continuum filters appear to represent nearly the same apparent size, with the 2.15 $\mu$m filter showing a slightly smaller apparent diameter.  The 2.03 $\mu$m (H$_2$O) and 2.39 $\mu$m (CO plus H$_2$O) visibilities correspond to larger apparent sizes. 
   

\section{Data modeling}
\label{sec:model}
\begin{figure*}[t]
\hbox{
   \includegraphics[bb=60 58 565 775 , angle=90, width=8.95cm]{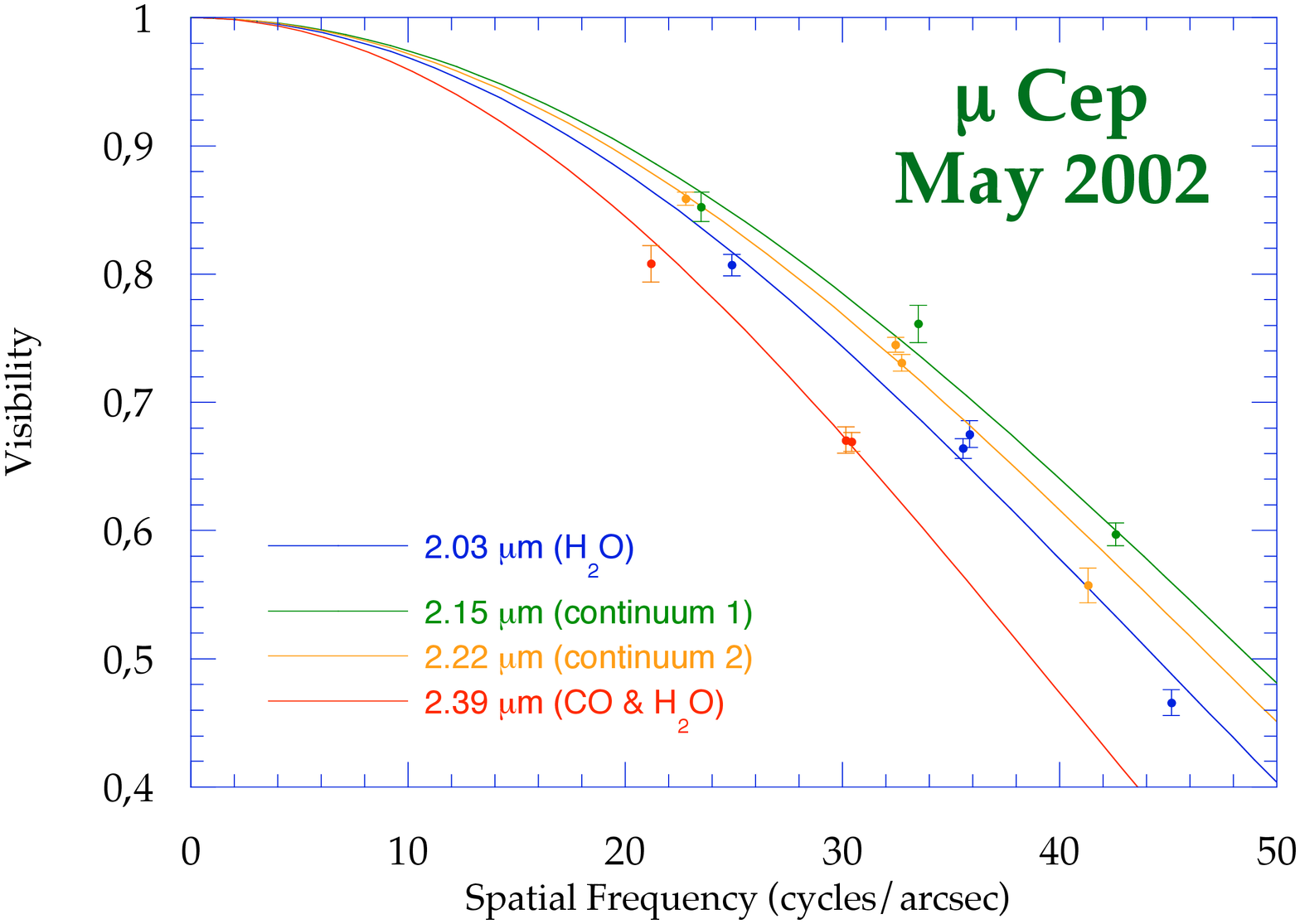}
   \includegraphics[bb=60 58 565 775 , angle=90, width=8.95cm]{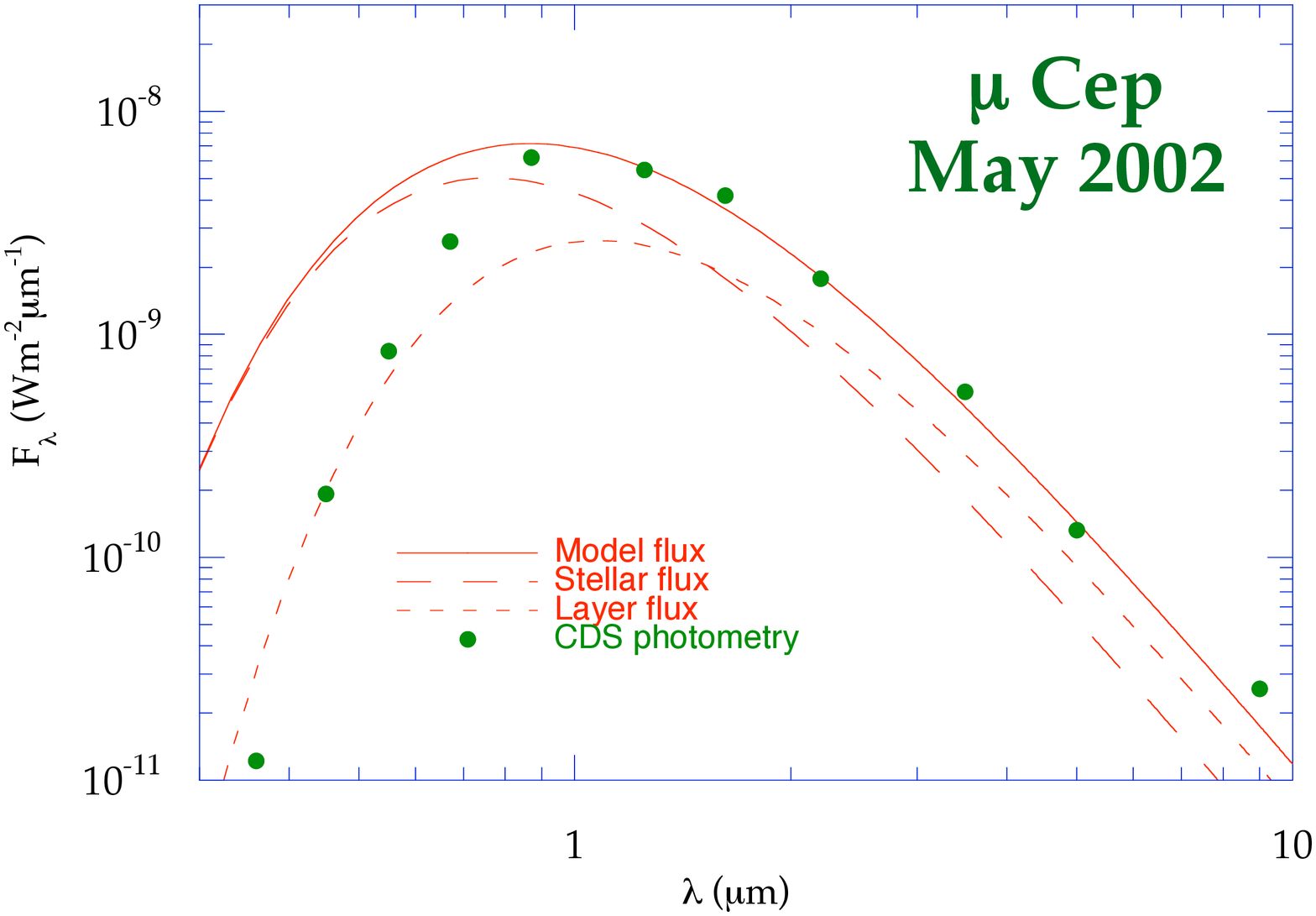}
}   
     \caption{Visibilities of $\mu$~Cep observed in narrow bands and fitted by the photosphere + spherical layer model (left). Photometry from the Simbad database compared to the synthetic spectrum derived from the visibility model (right).}
         \label{fig:mu_Cep}
   \end{figure*}

As Fig.~\ref{fig:mu_Cep} shows, the apparent diameter of the star is found bigger in the CO and H$_2$O bands than in the continuum bands. We found the same behavior for Mira stars. In \cite{perrin2004b}, the visibilities of Mira stars were successfully explained with a simple model. A photosphere of diameter $\O_{\star}$ emits as a blackbody of temperature 
$T_{\star}$.  It is surrounded by a spherical layer of diameter 
$\O_{\mathrm{layer}}$ with no geometrical thickness.  The layer is 
characterized by its temperature $T_{\mathrm{layer}}$ and its optical 
depth $\tau_{\lambda}$.  The layer absorbs the radiation emitted by 
the star and re-emits like a blackbody.  Scattering is neglected for 
our range of wavelengths.  The space between the layer and the stellar 
photosphere is empty. Intrinsic limb darkening for the photosphere is neglected, but the composite model has interesting brightness variations, more complex than classical limb darkening.

The analytical expression of the model is given by the equations below:
\begin{eqnarray}
I(\lambda,\theta) & = & 
B(\lambda,T_{\star})\exp(-\tau(\lambda)/\cos(\theta)) \\ \nonumber 
& & +B(\lambda,T_{\mathrm{layer}})\left[1-\exp(-\tau(\lambda)/\cos(\theta))\right]
\end{eqnarray}
for $\sin(\theta) \leq \O_{\star}/\O_{\mathrm{layer}}$
and:
\begin{equation}
I(\lambda,\theta)=B(\lambda,T_{\mathrm{layer}})\left[1-\exp(-2\tau(\lambda)/\cos(\theta))\right]
\end{equation}
otherwise, where $B(\lambda,T)$ is the Planck function and $\theta$ is 
the angle between the radius vector and the line between the observer 
and the center of the star. The intensity distribution is calculcated at 
the central wavelength of the filter. The effect of bandwidth has been 
neglected as the maximum error on the visibility model is less than 
0.4\% in the bands we have used (\cite{perrin2004c}). The visibility model is obtained by 
taking the Hankel transform of the circularly symmetric 
intensity distribution.  Although it is a quite 
simple view of the atmosphere of an evolved object, it is very 
convenient to use as it only depends on a small number of parameters 
and allows relatively easy and quick computations. It is important to 
note that the only physical parameters which vary from one wavelength 
to another are the optical depths. The temperature and geometrical 
parameters are the same at all wavelengths. 

   
\begin{table*}[htbp]
\caption[]{Log of observations.}
\label{tab:obs}
\begin{center}
\begin{tabular}{cccccll}
\hline
\hline
\noalign{\smallskip}
UT date & Filter         & Sp. Freq.               & Azimuth$^{\mathrm{a}}$   & $V^{2}$ & Calibrator 1 & Calibrator 2\\
               &  & (arcsec$^{-1}$)    & ($\degr$) &                &  & \\
\noalign{\smallskip}
\hline
\noalign{\smallskip}
5/31/2002 & K203 & 45.15 & 89.46 & $0.2171\pm0.0094$ & HD198149 & \\
6/4/2002 & K203 & 35.85 & 108.09 & $0.4560\pm0.0140$ & HD197989 & HD187076\\
6/5/2002 & K203 & 24.91 & 97.70 & $0.6515\pm0.0135$ & HD187076 & \\
6/6/2002 & K203 & 35.56 & 113.37 & $0.4411\pm0.0102$ & HD187076 & HD206778\\
\\
5/31/2002 & K215 & 42.59 & 91.12 & $0.3565\pm0.0106$ & HD130144 & \\
6/5/2002 & K215 & 23.49 & 98.30 & $0.7268\pm0.0196$ & HD187076 & \\
6/6/2002 & K215 & 33.50 & 114.21 & $0.5796\pm0.0221$ & HD187076 & HD206778\\
\\
5/31/2002 & K222 & 41.32 & 91.47 & $0.3107\pm0.0149$ & HD130144 & \\
6/4/2002 & K222 & 32.74 & 109.86 & $0.5342\pm0.0093$ & HD197989 & HD187076\\
6/5/2002 & K222 & 22.78 & 98.91 & $0.7375\pm0.0087$ & HD187076 & HD206778\\
6/6/2002 & K222 & 32.46 & 114.78 & $0.5551\pm0.0086$ & HD187076 & HD206778\\
\\
6/4/2002 & K239 & 30.43 & 110.78 & $0.4478\pm0.0099$ & HD197989 & HD187076\\
6/5/2002 & K239 & 21.19 & 99.67 & $0.6533\pm0.0230$ & HD187076 & \\
6/6/2002 & K239 & 30.17 & 115.54 & $0.4497\pm0.0139$ & HD187076 & \\
\noalign{\smallskip}
\hline
\end{tabular}
\end{center}
\begin{list}{}{}
\item[$^{\mathrm{a}}\,\,$counted positive from East towards North]
\end{list}
\end{table*}


We have used this model to fit the $\mu$~Cep data. The resulting fits are shown in Fig.~\ref{fig:mu_Cep}. The model can easily reproduce the chromatic dependence of the diameter, and the parameters of the model are strongly determined. The deviations of the observed visibilities from the model are generally consistent with the errors, except for the highest spatial frequency visibility in the 2.22 $\mu$m band, which is low by about 3$\sigma$.  This could reflect an inadequacy of the model, which surely over-simplifies the actual physical situation. This particular point will be examined in Sect.~\ref{sec:MarkIII}.

The procedure to find the best set of parameters has been described in \cite{perrin2004b}. The best parameters of the fit are listed in Table~\ref{tab:mucep}. The optical depth of the spherical layer is almost negligible in the continuum bands. It is larger in the H$_2$O band in K203, although rather small, and quite large in the CO and H$_2$O bands in K239.

As in the model fitting of Mira stars, the temperatures are not well determined by the visibility data alone. The $\chi^2$ surface has the shape of a trough in the $(T_{\star}, T_{\mathrm{layer}})$ plane as the visibilities only constrain the relative brightness of the photosphere and of the spherical layer. It has been necessary to impose a photometric constraint. We therefore have forced the flux emitted in the K band by the model to be consistent with the K band magnitude of the star. No contemporaneous magnitude was available. We calculated a mean value and a precision from the values listed in the CIO catalog (Gezari et al. 1993): $\mathrm{K}=-1.75\pm0.10$. The error bars on temperatures have been obtained by measuring the range of temperatures at minimum $\chi^2$ only. In the case of $\mu$~Cep, the $\left[ \chi^2 , \chi^2 + 1 \right]$ interval covers a large range of temperatures, especially for $T_{\star}$. The large $T_{\star}$ values are compensated by large $\tau_{2.39}$ to keep consistency with the K magnitude range. Such a degeneracy could be avoided if visibilities at larger spatial frequency in K239 were available: the shape of the visibility curves is not the same for optically thin and optically thick layers (see Fig.~2). Our error bars on temperatures are therefore underestimated but providing better error estimates from our current data set is unfortunately not possible. 

We have extrapolated our model to other wavelength ranges to compute a spectral energy distribution (Fig.~\ref{fig:mu_Cep}). Since our model is not a spectroscopic model, fluxes cannot be assessed in bands other than K unless we assume ad-hoc optical depths. We have chosen a constant optical depth of 0.70 which lies in the range of the optical depths measured in the narrow bands and accounts for the total observed flux in K. Absorption by molecules such as TiO at short wavelengths being not taken into account here, the graph is only indicative of the shape of the spectrum in the near-infrared. Absorption by TiO is negligible in that range. UBVR fluxes are from the Simbad data base. Infrared fluxes are averages of the values of the CIO catalog. This synthetic SED is only indicative of the global shape of the spectrum based on our K band measurements. Near-infrared fluxes are well reproduced by the model. Above 5~$\mu$m the model underestimates the flux as the excess due to dust is not taken into account here. At shorter wavelengths, the flux is attenuated by the molecular absorption, primarily TiO. 


\begin{table}[htbp]
\caption[]{Best model fit parameters for $\mu$~Cep.}
\label{tab:mucep}
\begin{center}
\begin{tabular}{ll}
\hline
\hline
\noalign{\smallskip}
Year & 2001 \\
\noalign{\smallskip}
\hline
\noalign{\smallskip}
$\O_{\star}$ (mas) & $14.11\pm0.60$ \\
$\O_{\mathrm{layer}}$ (mas) & $18.56\pm0.26$ \\
$T_{\star}$ (K) & $3789\pm100$ \\
$T_{\mathrm{layer}}$ (K) & $2684\pm100$ \\
$\tau_{2.03}$ & $0.22\pm0.03$ \\
$\tau_{2.15}$ & $0.02\pm0.01$ \\
$\tau_{2.22}$ & $0.07\pm0.01$ \\
$\tau_{2.39}$ & $3.92\pm1.58$ \\
\noalign{\smallskip}
\hline
\end{tabular}
\end{center}
\end{table}


\section{Bolometric flux and effective temperature}
\label{sec:parameters}

Before estimating the bolometric flux of the source, it is necessary to assess the amount of extinction and the sources of extinction. Using the intrinsic colors of \cite{elias1985} for M2~Ia supergiants we derive an average intrinsic color $(B-V)_0=1.70$. From the Simbad magnitudes, this yields a color excess $E_{B-V}=0.56$ for $\mu$~Cep. The reddening law of \cite{mathis90} for diffuse interstellar dust yields an extinction $A_v=1.74$. Using the law of \cite{elias1985} for circumstellar dust in supergiants would yield $A_v=1.99$. 

One may wonder what are the respective shares of circumstellar and interstellar reddening and this raises the question of the distance of $\mu$~Cep. Several parallax measurements or distance estimates are available in the literature (Perryman et al. 1997, Yuasa et al. 1999,  Humphreys 1978) with values in the range 500-1600 pc and large uncertainties.  Such a large range is not a sufficient constraint on the extinction. We propose to work back from the angular measurements.  Since $\alpha$~Ori and $\mu$~Cep have similar spectral types, one may expect the characteristics of these stars to be close. In particular the ratios between the photospheric diameter and the distance of the molecular layer from the photosphere found in \cite{perrin2004a} for $\alpha$ Ori and in this paper  for  $\mu$~Cep are in both cases approximately 3.  Assuming the parallax of Betelgeuse is correct, the scaling of diameters requires a distance of 390~pc for $\mu$~Cep in agreement with the parallax of \cite{humphreys1978}.  We adopt as an error, the difference with the estimate of \cite{humphreys1978}, about 140~pc. We would get the same error by scaling Betelgeuse's parallax and keeping the original error on the parallax.  Following this method and given that no accurate distance estimate is otherwise available we adopt the following value:
\begin{equation}
d_{\mu~\mathrm{Cep}}=390\pm140~{\mathrm{pc}}
\end{equation}
Humphreys (1978)  found a visual extinction $A_ v$=2.41 for  $\mu$~Cep from intrinsic and observed colors. $\mu$~Cep belongs to the CEP~OB~2 association for which extinctions are found in the range $1.17-3.33$ for other supergiants. The lowest value can be understood as an upper limit of the extinction due to the interstellar medium.  \cite{perry1982} have measured interstellar extinction on stars within 300~pc from the Sun. Using their data yields $A_ v$$=0.3 - 0.4$ in the direction of $\mu$~Cep. \cite{neckel1980} have established extinction up to 3~kpc. The extinction law measured in field $\sharp 308$ to which $\mu$~Cep belongs clearly shows that extinction increases and reaches a plateau above 500~pc due to the presence of a cloud of interstellar dust. The value of extinction at the plateau is about 2.0 and the estimated visual extinction at 300~pc is consistent with the value extrapolated from the \cite{perry1982} data.  
Given this and the adopted $\mu$~Cep distance, we derive a visual extinction of $A_v=1.0$ from \cite{neckel1980} for the interstellar medium, in agreement with the upper limit of 1.17 derived from \cite{humphreys1978} and with the estimate of $1.13\pm0.42$ based on the model by \cite{arenou1992}. The consistency of this analysis shows that the distance we have estimated from a geometrical rationale is plausible. Using the \cite{mathis90} reddening law we deduce that 0.32 of the color excess is due to the interstellar medium and therefore 0.24 is due to circumstellar material.  Applying the \cite{elias1985} reddening law we obtain a total extinction $A_v=1.85$. 
 
 We have used the infrared data from the 1999 edition of the \cite{gezari93} catalog available at CDS to compute the bolometric flux. The data have been complemented by UBVR data from the Simbad database. Infrared data above 5\,$\mu$m have not been used as they are dominated by circumstellar dust emission. 
 The usual way to compute the bolometric flux is to integrate the discrete SED established from the available photometric data. This still underestimates the total flux. The SED is therefore extended by blackbody distributions towards short and long wavelength ranges.\\
 The assessment of the uncertainty on the bolometric flux is an issue with this method. We have adopted a different one. Photometric data are first corrected for interstellar and circumstellar reddening using the extinction laws of \cite{mathis90} for diffuse interstellar dust and of  \cite{elias1985} for supergiant intrinsic extinction. The de-reddened SED is now close to a blackbody distribution. It is fitted by a Planck function to derive the total emitted flux. In order to derive a realistic error bar on the bolometric flux, we have associated an ad-hoc common error bar to each photometric measurement by forcing the reduced $\chi^{2}$ to 1. This technique accounts for the average measurement uncertainty and for the photometric variability of the object 
as the measurements span different epochs. Two parameters are constrained by the fitting procedure: a scaling factor proportional to the total flux and a temperature. Errors computed on these two parameters are used to derive an error on the bolometric flux by a Monte-Carlo method. Gaussian distributions are generated for the two parameters to produce a statistical distribution of the bolometric flux and derive its uncertainty. This yields a bolometric flux of $(17.62\pm3.44) \times 10^{-13}\,{\mbox{Wcm}}^{-2}$. It is to be compared to the value adopted by \cite{mozurkewich2003} of $14.46 \times 10^{-13}\,{\mbox{Wcm}}^{-2}$ computed with the integration method. The two values are compatible given the error bar we have derived despite the use of very different methods. \cite{mozurkewich2003} have corrected fluxes for interstellar redden-
ing assuming that the amount of flux absorbed by the circumstellar
environment is re-radiated at longer wavelength and therefore contributes
to the integrated flux provided the flux data set is complete. The smaller
value may be due to either an uncomplete sampling of the SED in their analysis, or to an
overestimation of circumstellar reddening in our method. 

The effective temperature is determined by assuming that the star 
radiates as a blackbody and has a physical diameter given by the measured photosphere diameter.  The effective temperature is then:

\begin{equation}
    {\mbox{T$_{\mbox{\scriptsize eff}}$}}=7400\,{\left( 
    \frac{{\mbox{F$_{\mbox{\scriptsize bol}}$}}}{10^{-13}\,{\mbox{Wcm}}^{-2}} 
    \right)}^{1/4}{\left( 
    \frac{1\,{\mbox{mas}}}{\O_{\star}}\right)}^{1/2}\,\,{\mbox{K}}
    \label{eq:teff}
\end{equation}
which yields $4024\pm221$~K for $\mu$~Cep (the uncertainty is computed by a Monte-Carlo method). The effective temperature is 235~K larger than the temperature provided by the direct fit of  the visibilities but is consistent within the error bars. Contrary to the systematically lower temperatures found for $\alpha$~Her and $\alpha$~Ori compared to the temperature of a giant with the same spectral type in \cite{perrin2004a}, the effective temperature of $\mu$~Cep is found larger (the effective temperature of an M2 giant is 3730~K according to \cite{ridgway1980}). We can suggest two possible explanations. First, there may be an issue with the flux estimate given current data sets as we note that the model temperature is found consistent with the spectral type of $\mu$~Cep. If the temperature difference can be considered significant, a more interesting possibility is that the temperature is correct and the spectral type is not.  Spectral types in cool M stars are determined largely from molecular bands, and the strong bands will form in the upper atmosphere, which is characterized in our model by the layer temperature and optical thickness.  This upper layer is seen in absorption against the photosphere, and the apparent depth of bands is determined largely by the temperature differential.  We note that in  $\mu$~Cep both the derived photospheric temperature and the layer temperature are higher than for $\alpha$~Ori, so that there may be a tendency to have similar band depths.  More generally, these cool, luminous stars may not be adequately described by the conventional temperature-luminosity indices, but may additionally require a description of the non-hydrostatically supported molecular layer.   We do not suggest this possibility as a conclusion, but as a hypothesis to investigate, with more complete characterization of both stars with narrow band and thermal IR interferometry.

The determinations of photospheric and effective temperatures are not sufficiently precise to evaluate the possibility of backwarming of the photosphere, such as we detected in the Mira stars.  This will differ from star to star depending on the optical thickness of the layer.

The error on distance is unfortunately still very large and dominates uncertainties on absolute quantities. We derive estimates of the absolute luminosity and of the bolometric magnitude of $\mu$~Cep (the lower error is due to the uncertainties on flux and angular diameter, the upper error is due to the adopted uncertainty on distance):
\begin{eqnarray}
L_{\mathrm{\mu~Cep}}          & = & 9.5_{\pm1.6}^{\pm6.2}~\times 10^{4} ~L_{\odot} \\ \nonumber
M_{\mathrm{bol, \mu~Cep}} & = & -7.4_{\pm0.2}^{ \pm1.1} \\ \nonumber
\end{eqnarray}
From \cite{allen2000} a bolometric magnitude of -7.22 is expected for a supergiant of type M2 in agreement with our result. 

   
\begin{figure*}[t]
\hbox{
   \includegraphics[bb=60 58 565 775 , angle=-90, width=8.9cm]{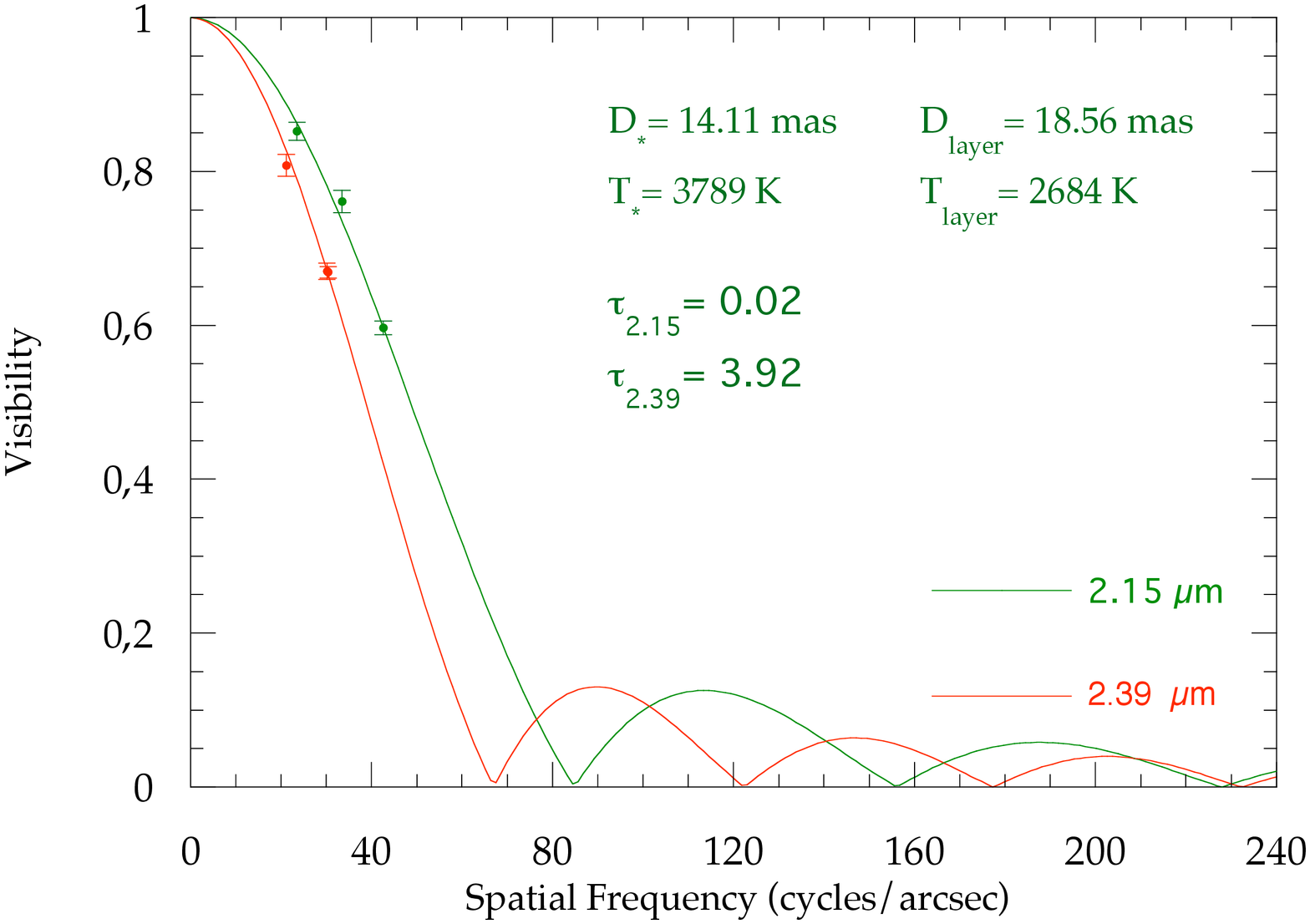}
   \includegraphics[bb=60 58 565 775 , angle=-90, width=8.9cm]{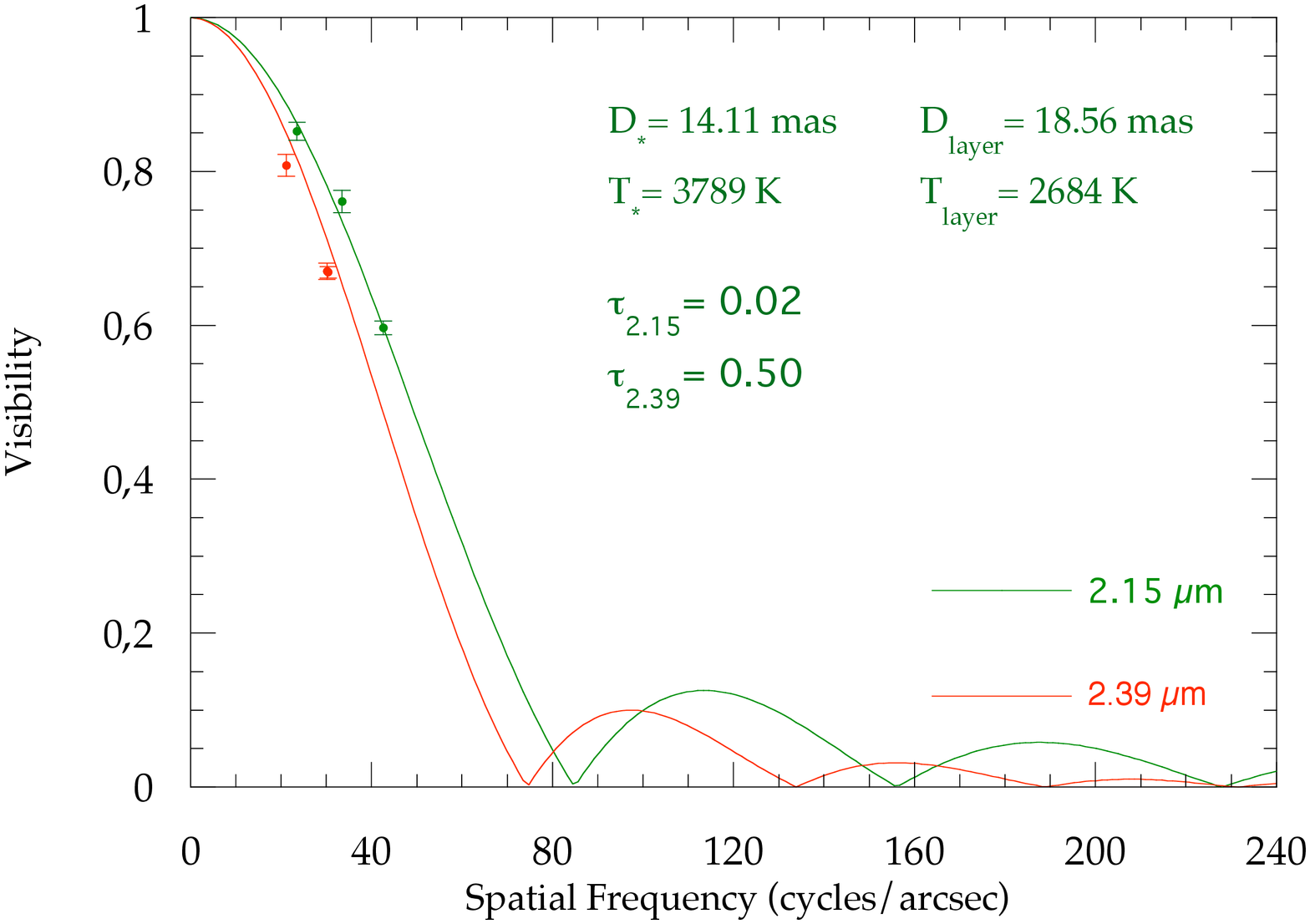}
}   
\vspace{0.4cm}
\hbox{
   \includegraphics[bb=60 58 565 775 , angle=-90, width=8.9cm]{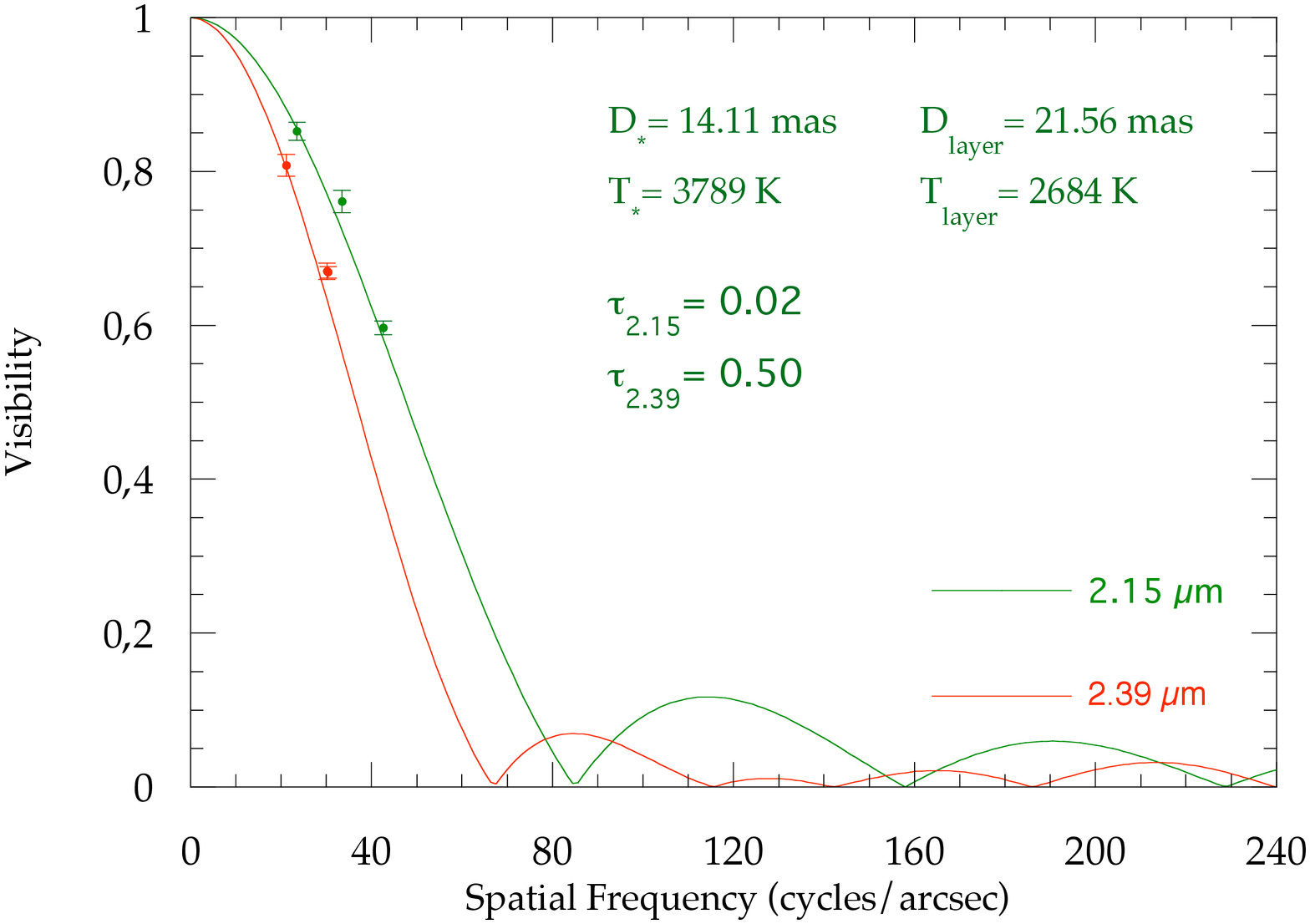}
   \includegraphics[bb=60 58 565 775 , angle=-90, width=8.9cm]{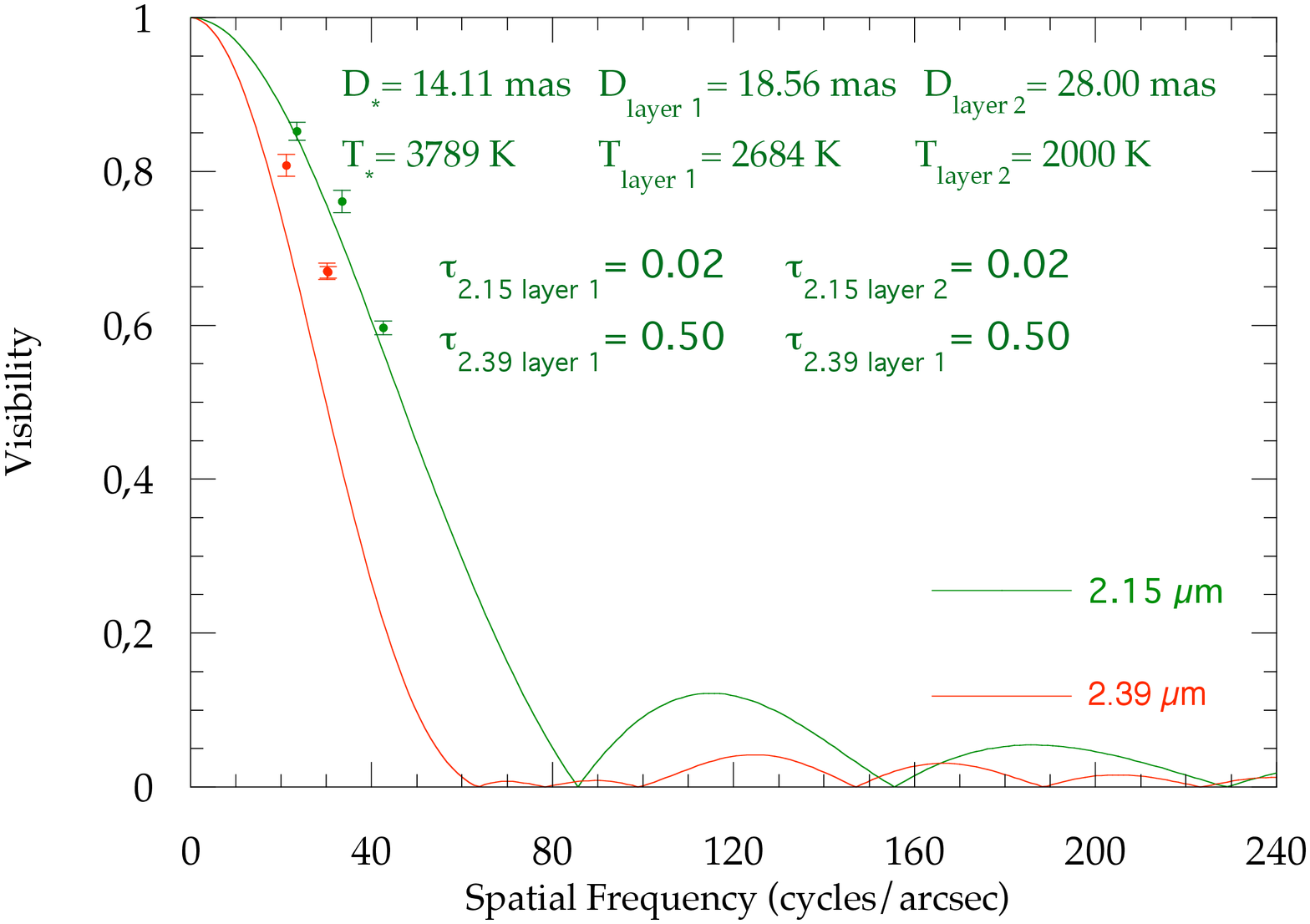}
}   
     \caption{Simulation of visibilities for $\mu$~Cep with different parameters. The best fit parameters found in this paper have been used for the top left model. Only simulated visibilities in the $2.15~\mu$m and $2.39~\mu$m filters are plotted here. The visibility measurements at these two wavelengths reported in this paper are also plotted}.
         \label{fig:modles}
   \end{figure*}


\section{Discussion}
\label{sec:discussion}
\subsection{Comparison with other interferometric measurements}
\label{sec:MarkIII}
$\mu$~Cep has been observed with the Mark III interferometer at 800~nm (Mozurkewich et al. 2003). The Mark III sampled almost exactly the same spatial frequencies as ours: 18.2, 26.1, 32.1 and 49.7 arcsec$^{-1}$. The measured diameter is $18.672\pm0.435$~mas. But the authors had to model the object with two components: a totally resolved component and the 18.7~mas disk. The resolved component causes the uniform disk visibility to drop at zero baseline from 1 to  $0.806\pm0.012$. The finding by Mark~III that an extra component is required can possibly explain why it is difficult to get a perfect fit of the higher frequency visibilities with our model. 

The comparison to our data is remarkable. The diameter found in the visible is equal to the diameter we find for the spherical layer. The same conclusion was reached for $\alpha$~Ori and for Mira stars. The molecular layer absorbs and radiates in the K band as a greybody (since it is partially transparent).  At shorter wavelengths, where molecules like TiO are strong absorbers, part of the absorbed radiation of the star is scattered by the layer (the cross section for Rayleigh scattering is about 80 times larger at 800~nm than it is in the K band) while the other part is absorbed and thermally radiated by the TiO molecules. The visibility function at 800~nm should therefore depart from a uniform disk and be somewhat Gaussian. This would appear mainly at higher spatial frequencies, but the baselines used by Mark III for the observations were too short as  the minimum observed visibility was about 40\%. At lower spatial frequencies the difference between the two functions is too small to be detected.

With this interpretation, Mark~III has measured the size of the molecular layer around $\mu$~Cep. This explains why they find a cooler effective temperature ($3181\pm52$~K). The comparison of the two sets of measurements shows that both obtain the same size for the molecular layer,  hence the strong absorbers/scatterers at 800~nm are located in the same layer as CO and H$_2$O found from infrared measurements. \cite{mozurkewich2003} commented on the comparison of their short wavelength diameters with infrared measurements - cooler stars appear systematically smaller in the near-infrared than expected. We now understand that this is due to the presence of the molecular layer around the star. 

\subsection{Spectroscopic evidence of a molecular layer close to the photosphere}

$\mu$~Cep was observed in the late 60s with a balloon-borne infrared spectrograph - Stratoscope~II - which recorded spectra in wavelengths that are obscured by terrestrial water vapor.  \cite{danielson1965}Ê claimed for strong water-vapor bands. But later observations failed to confirm this and interpreted the balloon data  as due to CN (Wing \& Spinrad 1970). \cite{tsuji2000a} has reanalyzed the Stratoscope~II  spectra and interpreted the data (especially the 1.4 and 1.9~$\mu$m features) as due to the presence of warm H$_2$O ($1500\pm500$~K) above the photosphere. The same interpretation was reached for CO with Fourier Transform spectroscopy in \cite{tsuji1987}. Using ISO spectra taken with the Short-Wavelength Spectrograph, \cite{tsuji2000b} shows that features in emission can be explained by an envelope or shell of warm water located a stellar radius above the photosphere at the temperature he found with the shorter wavelength data.  Such an atmospheric component was not expected from classical stellar models, and Tsuji baptized this region the MOLsphere. 

The model of the MOLsphere is not exactly the same as ours. A more detailed investigation of this result is required here. The model of \cite{tsuji2000a} is also a simple one: an absorbing layer above the photosphere but with no constraint on its altitude above the photosphere. The relative shape of the spectrum has been fitted, but not its absolute flux. The model is therefore adequate to within a multiplicative factor. In \cite{tsuji2000b}, the temperature found by \cite{tsuji2000a} is used as a baseline and the distance of the layer is calculated to allow for water to appear in emission in the ISO spectra. As a consequence, an error on temperature in the first paper may induce an error on the distance of the layer. Since the modeling in the first paper was in relative flux, the geometrical characteristics of the layer may remain somewhat uncertain. \cite{tsuji2000a} chose a temperature between 1000 and 2000~K because it is compatible with the upper limit of 2800~K for the excitation temperature of the H$_2$O pure rotation lines by \cite{jennings1998} and accounts for the Stratoscope spectra. But a higher temperature of $2684\pm100$~K is still compatible with the upper limit and would considerably change the extension of the \cite{tsuji2000b} layer, emission being possible with a much more compact layer. At such a high temperature, emission by the layer is not negligible at 2~$\mu$m and could fill in the gap of the \cite{tsuji2000a} model above 2.3~$\mu$m in Fig.~3 of that paper.

One first possibility is therefore that the MOLsphere model should be re-examined with our new geometrical parameters, which may provide additional constraints on the difficult question of altitude. A second possibility is that two layers are present, one close to the photosphere and another one higher in the atmosphere at a cooler temperature. The cooler one would be much more difficult to see in our visibility data. An attempt to fit the data with a two layer model has failed: we cannot reproduce the low visibilities at 41-45~arcsec$^{-1}$ by adding a second cooler layer. We therefore may be just sensitive to the absorption of this cooler layer. From our measurement we have derived $\tau_{203}=0.20\pm0.03$ (cf. Table~\ref{tab:mucep}). From Fig.~3 of \cite{tsuji2000a} we read out a difference between the $log$ of the intrinsic photosphere flux  and of the measured flux at this wavelength of 0.1, hence a flux ratio of 1.26 equivalent to an optical depth of 0.23. This being consistent with our estimated optical depth, we believe the results are not contradictory. We therefore conclude that two possibilities must be considered. In the first case we may have a single layer close to the star, and in the second case, a more complex model may be required.  We suggest that for the next level of detail, two layers may be required, one close containing particularly CO and a higher one with water vapor.  As high spectral resolution interferometry of the CO bands becomes available (expected soon with the VLTI/AMBER instrument) it should be possible to constrain the depth dependence of CO temperature and density from the photosphere to the top of the molecular layer.

Recently, \cite{ohnaka2004} has examined the wavelength dependence of angular diameter and the H$_2$O spectra of $\alpha$~Ori and $\alpha$~Her, investigating a very simple, single layer model.  In the Ohnaka model, the molecular layer is modeled as an envelope of uniform density and temperature beginning at the photosphere and extending to a radius.  This model accounts for the characteristic stellar size variation with wavelength, explains the lack of strong spectral structure in the mid-IR, and the typical strength of spectral lines in the 6 $\mu$m band.  On the other hand, in a similar analysis of Mira stars \cite{ohnaka2004} found that it was necessary to adopt a model with two molecular layers above the photosphere.  Perhaps not surprisingly, the complexity of the required model may depend on both the importance of the molecular layer in a particular case (weaker in the supergiants and stronger in Mira stars), and the amount of information about the atmospheric structure available in the measurements (minimal in spectra and somewhat more in spatial visibilities).  

\subsection{Perspective for further interferometric studies of molecular layers}

The spectral and the spatial information are showing clearly that supergiants, as well as Mira stars, are able to maintain a layer of molecular material high above the photosphere.  While the phenomenon is more dramatic in Mira stars, the Miras present a particularly thorny observational problem.  Virtually every parameter is varying with time, and the variations are often not even repetitive from cycle to cycle.  The supergiant stars offer an interesting alternative.  The underlying star shows less variability, and a hydrostatic model, even a plane parallel model, can offer at least a zeroth order approximation for investigative purposes.  Further, the possibility of pursuing molecular layers in successively lower luminosity stars, hotter stars, etc... may shed some light on the key phenomena which enable the molecular layer support.

The availability of specific, albeit simple, empirical models such as described here and by \cite{ohnaka2004} greatly simplifies the problem of designing experiments which can clarify and refine the models themselves.  This is illustrated by Fig.~\ref{fig:modles}.  The upper left shows the computed visibility curves for the best-fit model discussed above.  A single parameter has been varied from one plot to the next in order to illustrate the sensitivity of the observations to the parameters. The model of the bottom right panel was computed with two layers. We have chosen to display the $2.15~\mu$m (continuum) and $2.39~\mu$m (CO and H$_2$O bands) models only for sake of clarity.

The impact of the layer optical depth is two-fold.  First, it changes the width of the first (central) visibility lobe, giving a direct indication of the changing apparent ``size'' of the star plus layer.  When the layer optical depth is very small, the source appears as just the photosphere (modeled here as a uniform disk, as is the case at $2.15~\mu$m in all panels), and when the layer optical depth is very large, the source appears as just the layer (which then also appears in this model as approximately a uniform disk, this is the case at $2.39~\mu$m in the upper left panel).  Thus in both extreme cases of high and low layer optical depth, the visibility appears in this model with strong higher-order visibility lobes.  However, for intermediate optical depths, the higher order lobes are reduced in amplitude (lower left panel at $2.39~\mu$m).  This is easy to understand, as the star+layer gives the appearance of a darkened disk, and it is well known that an impact of limb-darkening on the visibility distribution is to dampen the higher lobes.  Thus we see that the amplitudes of the higher lobes serve as a check on the star+layer model.  As the molecular layer is assumed to be more complex, e.g. two layers (lower right panel), this dampening of the higher lobes becomes even more significant.  Eventually for a geometrically very thick layer it may tend toward an approximate Gaussian shape (in both spatial brightness distribution and in visibility distribution).

These three examples show that the detailed form of the visibility curves are very sensitive to the stellar structure which may be easily constrained and refined with future observations, especially with longer baselines giving access to at least the second lobe of the visibility function.  Further, the actual limb darkening of the photosphere and of the layer can be detected with second lobe measurements in the very high and very low opacity filters.


\begin{acknowledgements}

The authors have made use of the photometric databases of the Centre de Donn\'ees Astronomique de Strasbourg (http://simbad.u-strasbg.fr). The authors are also very grateful to Dr. Keiichi Ohnaka, the referee, whose comments and questions helped to improve the quality of the paper. 

\end{acknowledgements}


\end{document}